# Moiré Engineering of Electronic Phenomena in Correlated Oxides

Xinzhong Chen[1‡], Xiaodong Fan[2‡], Lin Li[2‡*], Nan Zhang[2], Zhijing Niu[1], Tengfei Guo[3], Suheng Xu[1], Han Xu[2,4], Dongli Wang[2], Huayang Zhang[2], A. S. McLeod[5], Zhenlin Luo[4], Qingyou Lu[3], Andrew J. Millis[5,6], D. N. Basov[5]*, Mengkun Liu[1]*, Changgan Zeng[2]*

*[1]Department of Physics and Astronomy, Stony Brook University, Stony Brook, New York 11794, USA*

*[2]International Center for Quantum Design of Functional Materials, Hefei National Laboratory for Physical Sciences at the Microscale, CAS Key Laboratory of Strongly Coupled Quantum Matter Physics, Department of Physics, and Synergetic Innovation Center of Quantum Information & Quantum Physics, University of Science and Technology of China, Hefei, Anhui 230026, China*

*[3]High Magnetic Field Laboratory, Chinese Academy of Sciences, Hefei, Anhui 230031, China*

*[4]National Synchrotron Radiation Laboratory & CAS Key Laboratory of Materials for Energy Conversion, University of Science and Technology of China, Hefei, Anhui 230026, China*

*[5]Department of Physics, Columbia University, New York, New York 10027, USA*

*[6]Center for Computational Quantum Physics, The Flatiron Institute, 162 5th Avenue, New York, New York 10010, USA*

‡These authors contributed equally

*Corresponding authors: cgzeng@ustc.edu.cn, mengkun.liu@stonybrook.edu, lilin@ustc.edu.cn, db3056@columbia.edu

**Moiré engineering has recently emerged as a capable approach to control quantum phenomena in condensed matter systems[1–6]. In van der Waals heterostructures, moiré patterns can be formed by lattice misorientation between adjacent atomic layers, creating long range electronic order. To date, moiré engineering has been executed solely in stacked van der Waals multilayers. Herein, we describe our discovery of electronic moiré patterns in films of a prototypical magnetoresistive oxide $La_{0.67}Sr_{0.33}MnO_3$ (LSMO) epitaxially grown on $LaAlO_3$ (LAO) substrates. Using scanning probe nano-imaging, we observe microscopic moiré profiles attributed to the coexistence and interaction of two distinct incommensurate patterns of strain modulation within these films. The net effect is that both electronic conductivity and ferromagnetism of LSMO are modulated by periodic moiré textures extending over mesoscopic scales. Our work provides an entirely new route with potential to achieve spatially patterned electronic textures on demand in strained epitaxial materials.**

Moiré patterns emerge from overlaying two sets of mildly dissimilar periodic motifs. In two dimensional (2D) van der Waals systems, moiré patterns have been achieved by stacking atomic layers with slightly incommensurate periodicities and/or small twist angles[7]. Fascinating properties have emerged in 2D moiré systems including superconductivity and Mott insulating states in magic-angle graphene superlattices[1,2], moiré excitons in transition metal dichalcogenide heterobilayers[3–5], and topological conducting channels in twisted bilayer graphene[6]. However, moiré-type electronic modulations have never been experimentally demonstrated nor theoretically proposed in correlated transition metal oxides (CTMOs), which are known to host rich interactions between charge, spin, lattice, and orbital degrees of freedom[8,9]. Herein, we demonstrate that moiré engineering is applicable beyond van der Waals materials and report meso-scale spatial moiré modulations of coupled conductivity and ferromagnetism in strained thin films of a correlated oxide.

A universal property of CTMOs is the competition and coexistence of multiple order parameters that spontaneously give rise to spatially textured physical properties[10]. These emergent electronic textures are particularly prominent in manganites including $La_{1-x}Ca_xMnO_3$ and related materials[11–13]. However, deterministic creation of spatially ordered patterns "on-demand" remains a daunting task. Here, we utilize interactions between two distinct sources of periodic strain modulation in epitaxial LSMO films grown on LAO substrates. When these two modulations are collocated in real space, they coproduce moiré-like energy landscapes that locally modulate conductivity and ferromagnetism.

We investigated 20 nm thick LSMO films on LAO substrates (see Methods for details). Temperature-dependent resistivity and magnetic moment reveal a characteristic second-order phase transition from ferromagnetic metal (FMM) to paramagnetic metal (PMM) at $T_C \sim 340$ K (Fig. 1a). That $T_C$ is reduced compared to the bulk value ($\sim 370$ K[14]) attests to the large in-plane compressive strain imposed by the substrate[15]. We harnessed epitaxial control to define the local compressive strain patterns ultimately giving rise to moiré textures. First, we note that LSMO films on LAO possess highly oriented rhombohedral twin domains[16–20] that stem from local relaxations of substrate-imposed shear strain (see Fig. 1b, Extended Data Fig. 3 and Supplementary Note 2). As schematically shown in Fig. 1c, such a periodic structure results in a striped strain modulation[17,20], which we designate as domain stripes (DS). On the other hand, surface miscut steps in the LAO substrate (see Figs. 1d, e, and Extended Data Fig. 4) serve as another source for periodically modulated strain in LSMO. The resultant periodic strain modulation in LSMO, which we name as miscut stripes (MS), does not necessarily align to any specific crystal axis but depends strictly on the local miscut surface of the substrate[21] (see Supplementary Note 3). The spatial coexistence of DS and MS modulations yields moiré landscapes with domain sizes governed by the periodicities and relative angles of the two constituent stripe motifs (Fig.1f).

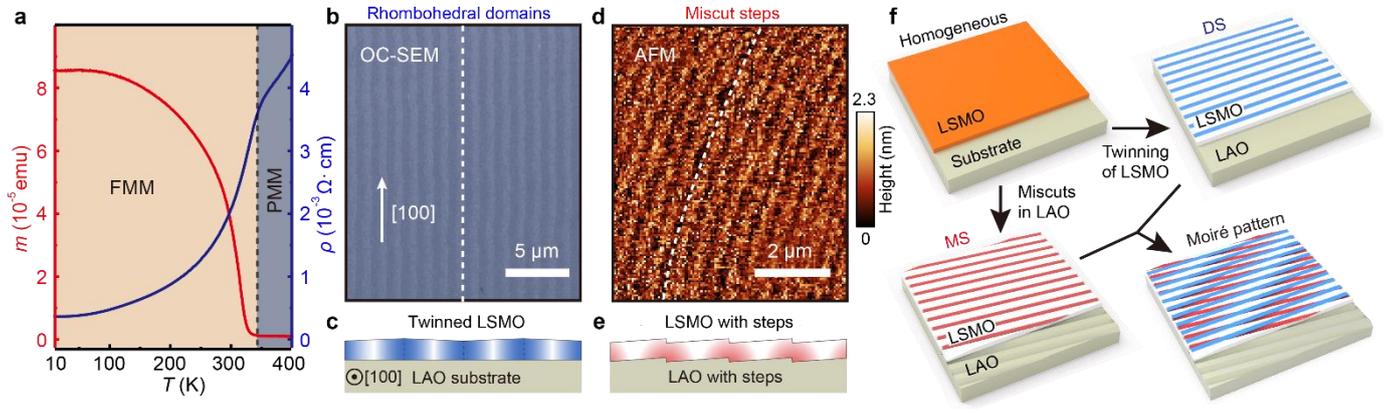

**Figure 1 | Two types of periodic strain modulations in LSMO thin films. a,** Temperature-dependent resistivity and magnetic moment of an LSMO thin film under study. The magnetic moment was measured under 100 Oe on warming after cooling the sample in a 100 Oe field. The black dashed line indicates $T_C$. **b,** Orientation-contrast scanning electron micrograph (OC-SEM) image of a LSMO thin film, showing surface rhombohedral domains with a width of ~590 nm aligned along the [100] direction. **c,** Schematic of the rhombohedral domain-induced strain modulation in a LSMO-LAO cross-section. **d,** Atomic force microscopy (AFM) image of a LSMO thin film, showing topographic surface changes due to crystal miscut steps in the LAO substrate. The period is ~500 nm. **e,** Schematic of the miscut step-induced strain modulation in a LSMO-LAO cross-section. White dashed lines in **b** and **d** indicate the directions of the rhombohedral domains and miscut steps, respectively. **f,** Schematic demonstration of how spatial coexistence of domain stripes (DS) and miscut stripes (MS) yields the combined moiré patterns comprising regions of local commensuration and incommensuration.

Nano-optical imaging provides direct evidence for the promise and possibility of moiré strain engineering. We utilized infrared (IR) scattering-type scanning near-field optical microscopy (s-SNOM) to visualize nano-scale moiré phenomena in strained LSMO/LAO (see Methods for detail). The amplitude of the nano-IR signal demodulated at the second harmonic of the tip tapping frequency ($S_2$) provides a local probe of the optical conductivity with ~25 nm spatial resolution[22,23]. In Fig. 2a, we display maps of the 2D nano-IR contrast, revealing periodic modulations of the local electronic response in LSMO/LAO. Notably, the periodicity $d$ is revealed to be considerably larger than that of either the DS or MS. Moreover, $d$ appears to be governed by the orientation of these patterns with respect to the [100] direction of LAO substrate (also see Extended Data Fig. 5). These systematic trends point a unified underlying mechanism for the formation of electronic patterns that we link to the emergent moiré pattern of combined periodic strains.

We now proceed to elucidate the rich real space structures of conductivity observed in our strained films. We first note that the two sets of overlaid unidirectional stripes with relatively small spacings readily produce longer period textures (top panels in Fig. 2b). Microscopic strain fields $\varepsilon_{DS}$ and $\varepsilon_{MS}$ that vary sinusoidally in space present a realistic scenario for our films. Here we associate the field $\varepsilon_{DS}$ with the domain stripes and $\varepsilon_{MS}$ with miscut stripes introduced in Fig. 1. As demonstrated in the bottom panels of Fig. 2b, we were able to reproduce the key trends in the images in Fig. 2a by assuming that the observed textures are a direct manifestation of the product $\varepsilon_{DS} \cdot \varepsilon_{MS}$ of these two sinusoidal strain fields. This product term naturally stems from the non-linear relationship between the strain and the conductivity, which roots in the strain-induced non-linear $T_C$ modulation in LSMO[15,24–26] (detailed in Supplementary Note 4). Our experimental result attests to the large magnitude of nonlinear coefficient for the cross term $\varepsilon_{DS} \cdot \varepsilon_{MS}$, which originates from the cubic (Oh) point symmetry of Mn in the cubic perovskite structure as well as the large Jahn-Teller effect due to strong electron-lattice coupling in LSMO2[24]. We note that in our simulated image, a local Gaussian average filter with radius ~100 nm is applied to account for the finite spatial resolution of the near-field probe and possible local strain relaxation (Extended Data Fig. 6).

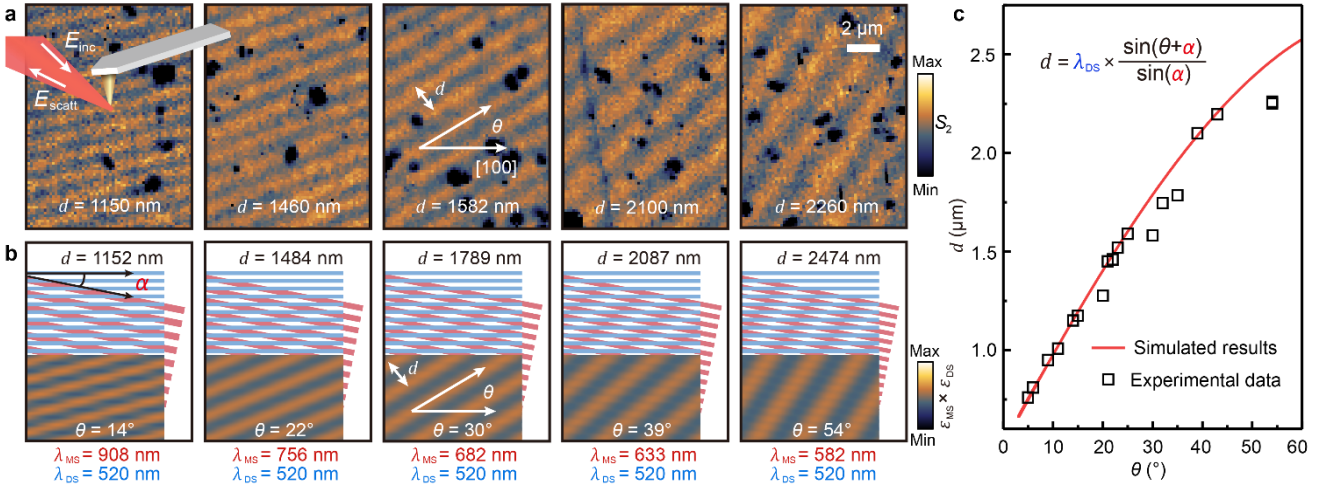

**Figure 2 | Electronic moiré patterns in LSMO films. a,** Nano-IR maps of the scattering amplitude $S_2$ obtained at the infrared wavelength of 10.7 μm and collected in different regions of a representative LSMO/LAO film. $d$ is the periodicity of textures that we attribute to moiré patterns, and $\theta$ is the angle between the orientation of the moiré textures and the [100] direction. **b,** Top: simulations of the moiré patterns formed by superimposing two periodic striped motifs, blue: DS, red: MS. From left to right, the periodicities of MS ($\lambda_{MS}$) are 908 nm, 756 nm, 682 nm, 633 nm, and 582 nm, respectively, while the periodicity of DS ($\lambda_{DS}$) and the twist angle between DS and MS are fixed at 520 nm and 11°, respectively. Bottom: simulations generated by multiplying the two striped patterns using the same parameters as in the upper half. **c,** The moiré patterns periodicity as a function of $\theta$. Black squares: experimental data. Red line: simulated results.

The concurrent evolution of the moiré fringe periodicity $d$ and angle $\theta$ (defined as the angle between the orientation of fringes and the LAO [100] direction) can be theoretically obtained by multiplying two sinusoidal waves with wavevectors $\boldsymbol{k}_{DS} = (0, \frac{2\pi}{\lambda_{DS}})$ and $\boldsymbol{k}_{MS} = (\frac{2\pi \sin(\alpha)}{\lambda_{MS}}, \frac{2\pi \cos(\alpha)}{\lambda_{MS}})$, where the $\lambda_{DS}$ and $\lambda_{MS}$ are the periodicities of DS and MS, respectively; and $\alpha$ is the twist angle between DS and MS. The resultant moiré pattern yields $\theta = \text{arccot}(\frac{\lambda_{MS} - \lambda_{DS} \cos(\alpha)}{\lambda_{DS} \sin(\alpha)})$ and $d = \frac{2\pi}{|\boldsymbol{k}_{DS} - \boldsymbol{k}_{MS}|} = \frac{\lambda_{DS} \sin(\theta + \alpha)}{\sin(\alpha)}$. We found the $d$-$\theta$ relation shown in Fig. 2a and Extended Data Fig. 5 can be reproduced by solely adjusting the $\lambda_{MS}$, whose gradual change is expected across the $1 \times 1$ mm² film due to a long-range variation in the surface miscut angle of the substrate and consequently in the MS periodicity. As shown in Figs. 2b, c, we found excellent agreement between experimental images and theory by monotonically varying $\lambda_{MS}$ from 908 nm to 582 nm while keeping $\lambda_{DS}$ and $\alpha$ constant. This agreement verifies that the spatial modulations in optical conductivity observed in the LSMO/LAO samples indeed arise from the synergistic effect of DS and MS. We note that similar moiré fringes were not found in the surface topography of LSMO co-recorded with the near-field imaging by AFM, which suggests that the large nonlinear coupling of optical conductivity to the local strain is an essential ingredient for deterministic occurrence of observed moiré patterns.

Apart from the unidirectional moiré fringes in Fig. 2, more complex moiré patterns can be obtained (Extended Data Fig. 8). For example, we observed curved moiré fringes in another LSMO/LAO sample of the same ~20 nm thickness (Fig. 3a). In Fig. 3b we show that such curved moiré patterns can also be perfectly reproduced by the product $\varepsilon_{DS} \cdot \varepsilon_{MS}$ under the assumption of MS with a slight curvature smaller than $\sim \frac{1}{100\,\mu m}$ (Supplementary Note 5). Thus obtained line profiles reveal an astonishing consistency between experiments and simulations, where the amplitude contrast of variations in optical conductivity resolved by our nano-imaging decays with decreasing periodicity (Fig. 3c, see also Extended Data Fig. 11). This consistency further confirms the validity of our interpretation. Furthermore, the variations of the orientation and curvature of the MS offer a unique opportunity to realize a diverse range of electronic moiré textures with different spatial configurations (Extended Data Fig. 9).

In Fig. 3d, another example of curved moiré pattern is presented. The near field line profile (Fig. 3f) demonstrates clear contrast between peaks and valleys of optical conductivity in the moiré pattern regions. On the other hand, regions without clear moiré pattern reveal optical conductivity approximately equal to the average of that of the moiré fringes. Note that the borders between moiré and non-moiré regions coincide with the twin boundaries of the LAO substrate (red dashed lines in Figs. 3d-f). In Fig. 3e, the topographic variation up to 10 nm in height comes from the LAO substrate twin structures[27,28] (Extended Data Fig. 2a and Supplementary Note 1). No visible moiré-like topographic modulation can be seen. Here we give a phenomenological interpretation: the miscut steps in LAO possess drastically different periodicities, orientations, and even curvatures on alternating sides of the underlying LAO twin boundaries (Extended Data Fig. 10). As simulated in Fig. 3g (detailed in Extended Data Fig. 9c), to the right of the LAO twin boundary, the small twist angle between DS and MS leads to a relatively large moiré pattern periodicity that can be observed under near-field imaging. To the left of the LAO twin boundary, the large twist angle is expected to produce a moiré pattern with a much smaller periodicity, on the same order of the DS or MS. As demonstrated in Fig. 3c and Extended Data Fig. 11, smaller periodicity yields a smaller IR near-field intensity contrast. The spatial features are therefore indistinguishable under the near-field mapping due to minimal conductivity variation between the electronic domains. This is supported experimentally by the average signal level detected for the non-moiré regions in Fig. 3f.

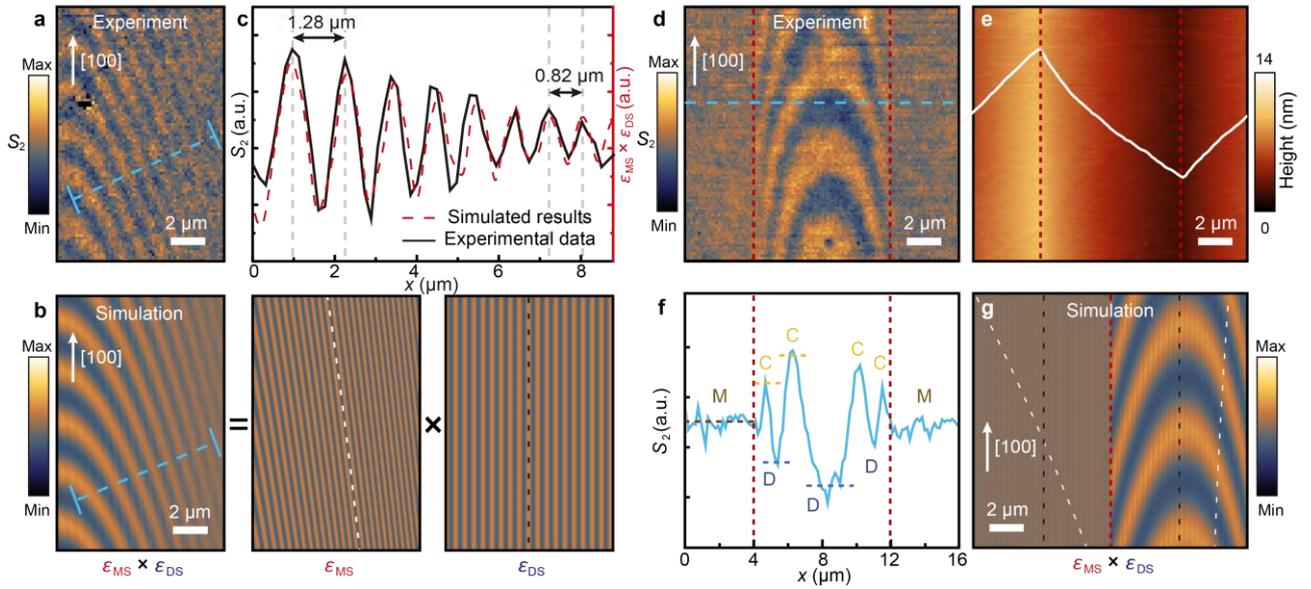

**Figure 3 | Curved electronic moiré patterns in LSMO films. a, b,** IR near-field image of a curved moiré pattern and the corresponding simulation, respectively. The simulation is generated by multiplying two periodic striped patterns representing MS and DS. The white and black dashed lines in **b** indicate the orientations of the MS and DS, respectively. **c,** Line profiles from the blue dashed lines in **a** and **b**, exhibiting high consistency between the experimental and the simulated contrast. **d, e,** IR near-field image and corresponding AFM image, respectively, showing alternating moiré and non-moiré regions across the LAO twin boundaries (indicated by red dashed lines). The white solid line in **e** is the AFM height profile. **f,** Line profile of the nano-IR contrast along the blue dashed line in **d**. The different signal levels are marked by "C", "D" and "M", which represent constructively strained, destructively strained and mixed strained region, respectively. **g,** Simulation of the image in **d** with the moiré pattern only visible on the right side. Note that the MS (orientation indicated by white dashed lines) changes orientation across the LAO twin boundary (red dashed line), while the DS (orientation indicated by black dashed lines) is consistently along the LAO [100] direction. The simulation details are shown in Extended Data Fig. 9c.

Temperature dependent s-SNOM and magnetic force microscopy (MFM) measurements provide further insight into the interdependence between electronic and magnetic moiré textures in strained LSMO. As a hallmark feature of the colossal magnetoresistive manganites, the microscopic mechanism of electrical conductivity in LSMO is intimately tied to the onset of long-range ferromagnetic order[9], and here we report moiré-like modulations of that dependency borne out and resolved at the meso-scale. Clear fringes in moiré conductivity are observed with a periodicity of ~1.05 μm at room temperature, as shown in the near-field images of Fig. 4a. Fig. 4b shows the MFM images acquired in roughly the same region, where clear frequency-shift contrast with the periodicity of ~1.13 μm is observed at 300 K, demonstrating the moiré modulations of local ferromagnetism. Increasing the temperature prompts a decrease in both the nano-IR contrast and magnetic response (Fig. 4c). This latter observation infers the positive correlation between the conductivity and the local ferromagnetic moment, consistent with the established phase diagram of LSMO[14]. The observed magnetic moiré fringes disappear at the ferromagnetic-paramagnetic transition temperature of 340 K (Fig. 4b and Extended Data Fig. 12), while the electronic counterpart persists with a greatly reduced nano-IR contrast up to all measurable temperatures (Fig. 4a). The temperature evolution of a curved moiré pattern is also shown in Fig. 4d. It is evident that the near-field optical contrast of the moiré pattern decays significantly at higher temperatures. Such observation further validates the electronic and magnetic properties of the moiré patterns are closely related to the $T_C$ of LSMO/LAO thin films.

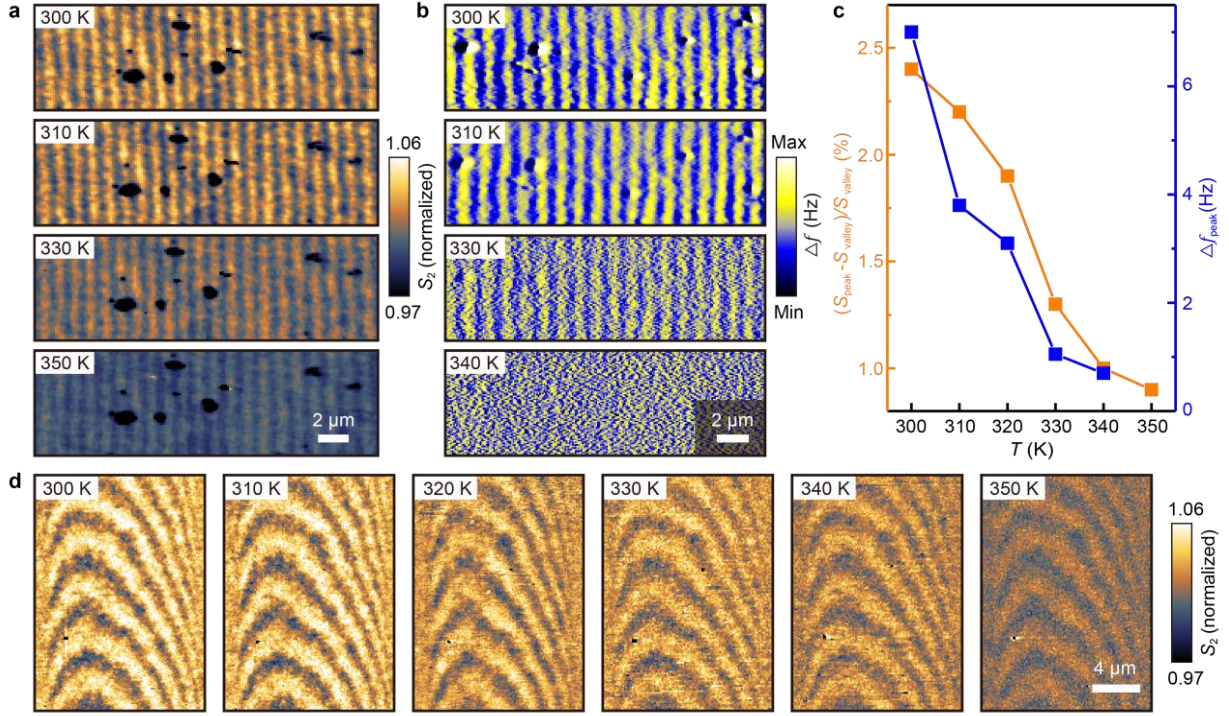

**Figure 4 | Temperature-dependent evolution of electronic and magnetic moiré patterns. a**, **b**, Near-field and MFM images of moiré pattern measured at elevated temperatures, respectively. **c**, Temperature dependent near-field contrast $(S_{peak} - S_{valley})/S_{valley}$ (orange) and MFM peak-frequency-shift $\Delta f_{peak}$ (blue) extracted from images in **a** and **b**, respectively. **d,** Near-field images of curved moiré pattern obtained at different temperatures. All of the near-field images are normalized to dark fringes to demonstrate the relative contrast fading with increasing temperature.

Lastly, we wish to point out that the individual DS and MS stripes are barely observable under s-SNOM, but their combined electronic effects nevertheless manifest clearly through the moiré pattern formation (Extended Data Fig. 13). This "magnified" electronic response may be potentially useful for identifying "hidden" textures of strain, electronic, or magnetic distribution in CTMO thin films, where individual traits are difficult to identify separately. Our novel approach also provides a strategy for manipulating complex electronic orders in strongly correlated quantum materials such as $NdNiO_3$[21,29] or $SrTiO_3$[30], where long-wavelength electronic and lattice modulations similar to those exploited here have been reported. Beyond the long-wavelength moiré patterns, it is also interesting to explore similar effects at shorter length scales. Materials with competing phases or proximate to a quantum phase transition can be tested with high priorities (for example the interplay between stripe and superconducting phases in high-$T_C$ cuprates is affected by strain).

**Methods**

**Sample fabrication and characterization.** LSMO thin films of varying thicknesses were grown on the LAO (001) substrates at 700 ℃ via pulsed laser deposition. A KrF excimer laser was used with an energy density of 5.5 J/cm$^3$ and a 3 Hz repetition rate. The O$_2$ pressure was maintained at 40 Pa during growth. The as-grown films were then in-situ annealed for 0.5 hour under the deposition conditions, followed by ex-situ annealing in a tube furnace at 780 °C in flowing O$_2$. The crystalline quality and lattice parameters of the films were characterized by X-ray diffraction (Panalytical X'Pert PRO MRD) at room temperature, with a typical result for the 20 nm sample shown in Extended Data Fig. 1. The magnetic properties were analyzed using a Quantum Design SQUID-VSM system. The electrical transport properties were measured using the four-probe method in a Quantum Design PPMS system.

**Infrared scattering-type scanning near-field optical microscopy (s-SNOM).** Measurements were conducted with two different s-SNOM apparatuses, yielding nearly identical results. Monochromatic light of wavelength ~11 μm (~10.7 μm for the other s-SNOM) was focused onto an atomic force microscope probe with tip apex radius ~25 nm. The AFM probe was operated in tapping mode with oscillation frequency Ω~250 kHz and amplitude A~50 nm (~80 nm for the other s-SNOM). Backscattered light from the tip-sample ensemble interfered with the reference light to yield phase resolved IR near-field detection. Lock-in demodulation at higher harmonics of Ω could greatly eliminate undesired background and preserve the genuine local near-field interaction emanating from the confined area below the tip apex.

**Magnetic force microscopy (MFM).** MFM measurements were performed under zero magnetic field by a home-made MFM. Using electron-beam deposition, the tip was coated with a trilayer film of 5 nm Ti, 50 nm Co, and 5 nm Au , which was subsequently magnetized with a permanent magnet along the tip axis. The MFM images were obtained in the frequency mode, in which the frequency shift $\Delta f$ ($\Delta f = f - f_0$) was "tracked" by a commercial phase locked loop. A negative frequency shift indicates an attractive tip−sample interaction ($\Delta f < 0$), while a positive frequency shift indicates a repulsive tip−sample interaction ($\Delta f > 0$).

**Orientation-contrast scanning electron micrograph (OC-SEM).** The rhombohedral domains in the LSMO/LAO samples were characterized using OC-SEM imaging, with misorientation between different domains represented by different contrast levels. The measurements were conducted in a field emission scanning electron microscope (ZEISS GeminiSEM 500) using the in-lens mode. The typical values of the acceleration voltage, objective aperture size, and working distance were 1 kV, 120 μm, and 5 mm, respectively.

**Data availability**

The datasets generated and analyzed during the current study are available from the corresponding author on reasonable request.

**Supplementary Information** is linked to the online version of the paper.

**Acknowledgements** This work was supported in part by the National Natural Science Foundation of China (Grants No. 11434009, 11804326, U1832151, and 11675179), National Key Research and Development Program of China (Grants No. 2017YFA0403600), Anhui Initiative in Quantum Information Technologies (Grant No. AHY170000), Hefei Science Center CAS (Grant No. 2018HSC-UE014), Anhui Provincial Natural Science Foundation (Grant No. 1708085QA20), and the Fundamental Research Funds for the Central Universities (Grant No. WK2030040087). This work was partially carried out at the USTC Center for Micro and Nanoscale Research and Fabrication.

**Author contributions** M.L., C.Z. and D.N.B. designed and supervised the work. X.C. and X.F. performed the s-SNOM measurements with assistance from Z.N., S.X., D.W., and H.Z.; L.L., N.Z., and H.X. fabricated the samples and performed the XRD, magnetic, and transport characterizations; T.G. and Q.L. performed the MFM measurements; X.C., L.L., X.F., D.N.B., M.L., and C.Z analyzed the data and wrote the manuscript. A.S.M., A.J.M. and Z.L. contributed to data interpretation and presentation. All authors contributed to the scientific discussion and manuscript revisions.

**Author information** The authors declare no competing financial interests. Correspondence and requests for materials should be addressed to C.Z. (cgzeng@ustc.edu.cn), M.L. (mengkun.liu@stonybrook.edu), L.L. (lilin@ustc.edu.cn), or D.N.B. (db3056@columbia.edu).

**Extended data**

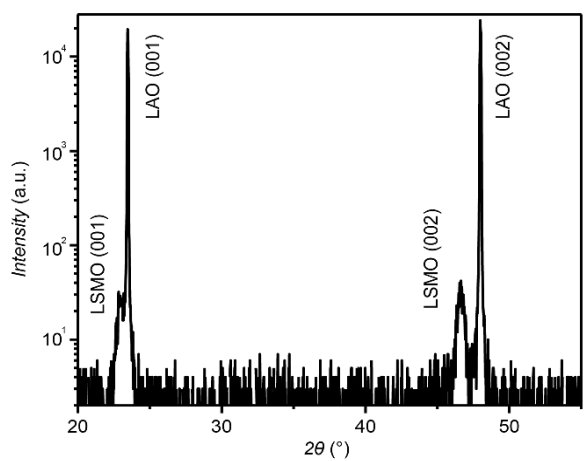

**Extended Data Figure 1 | X-ray diffraction spectrum of a 20-nm-thick LSMO thin film grown on LAO substrate.**
The LSMO thin films are clearly single-crystalline and coherently epitaxial on the LAO (001) substrates. The LSMO film's out-of-plane lattice constant as determined by the XRD peak is c=3.90 Å, which slightly exceeds the bulk value of 3.87 Å. The out-of-plane expansion, or tensile strain, proceeds from the compressive in-plane strain from the LAO substrates (pseudo-cubic structure with a lattice constant of 3.79 Å).

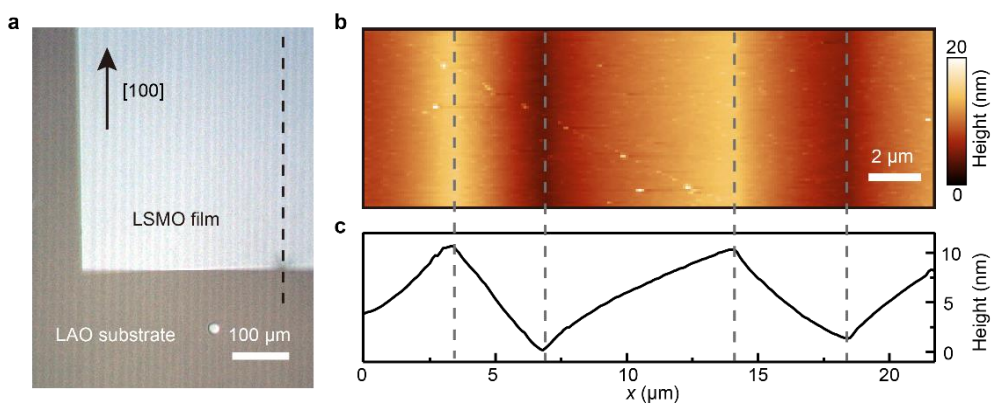

**Extended Data Figure 2 | Twin structures in LAO substrates. a,** Optical image of a LAO substrate with the surface partially covered by LSMO thin films. Twin structure fringes of the LAO substrate are visible across the LSMO film edge, lying parallel to the guiding black dashed line. **b, c,** AFM image and corresponding height profile of the twin structures in the LAO substrate.

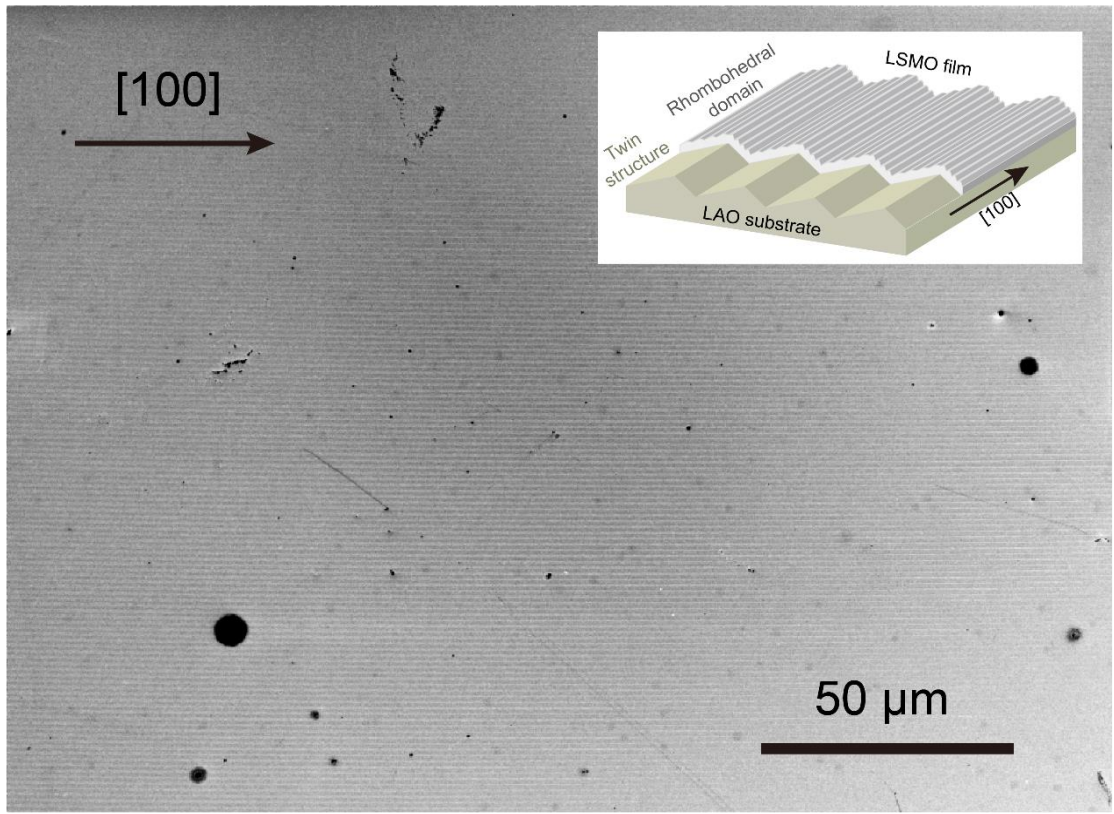

**Extended Data Figure 3 | Rhombohedral domains in the LSMO films on LAO substrates.** Large-scale OC-SEM image of a LSMO thin film showing surface rhombohedral domains. The inset demonstrates that the direction of rhombohedral domains in LSMO is identical to that of the LAO twins, running along the LAO [100] direction.

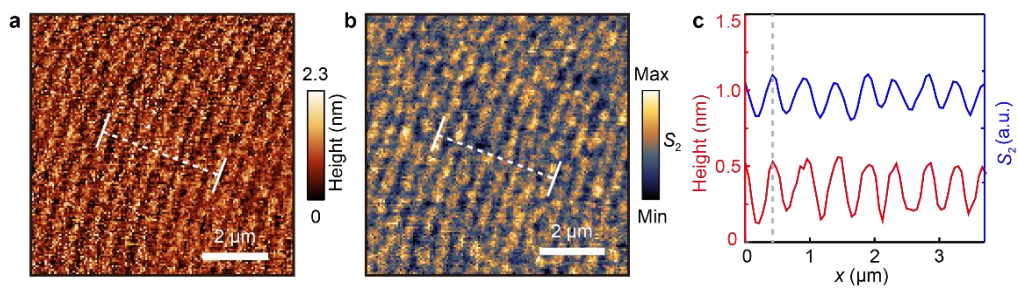

**Extended Data Figure 4 | Miscut stripes (MS)-induced electronic pattern in LSMO. a,** AFM image of a LSMO/LAO sample showing surface steps due to miscuts on the LAO substrate. **b,** Corresponding near-field image obtained simultaneously with the AFM image in **a**, showing spatially alternating values of IR near-field signal. **c,** Line profiles along the dashed lines in **a** and **b** show that near-field signal is positively correlated to the topography of miscut steps.

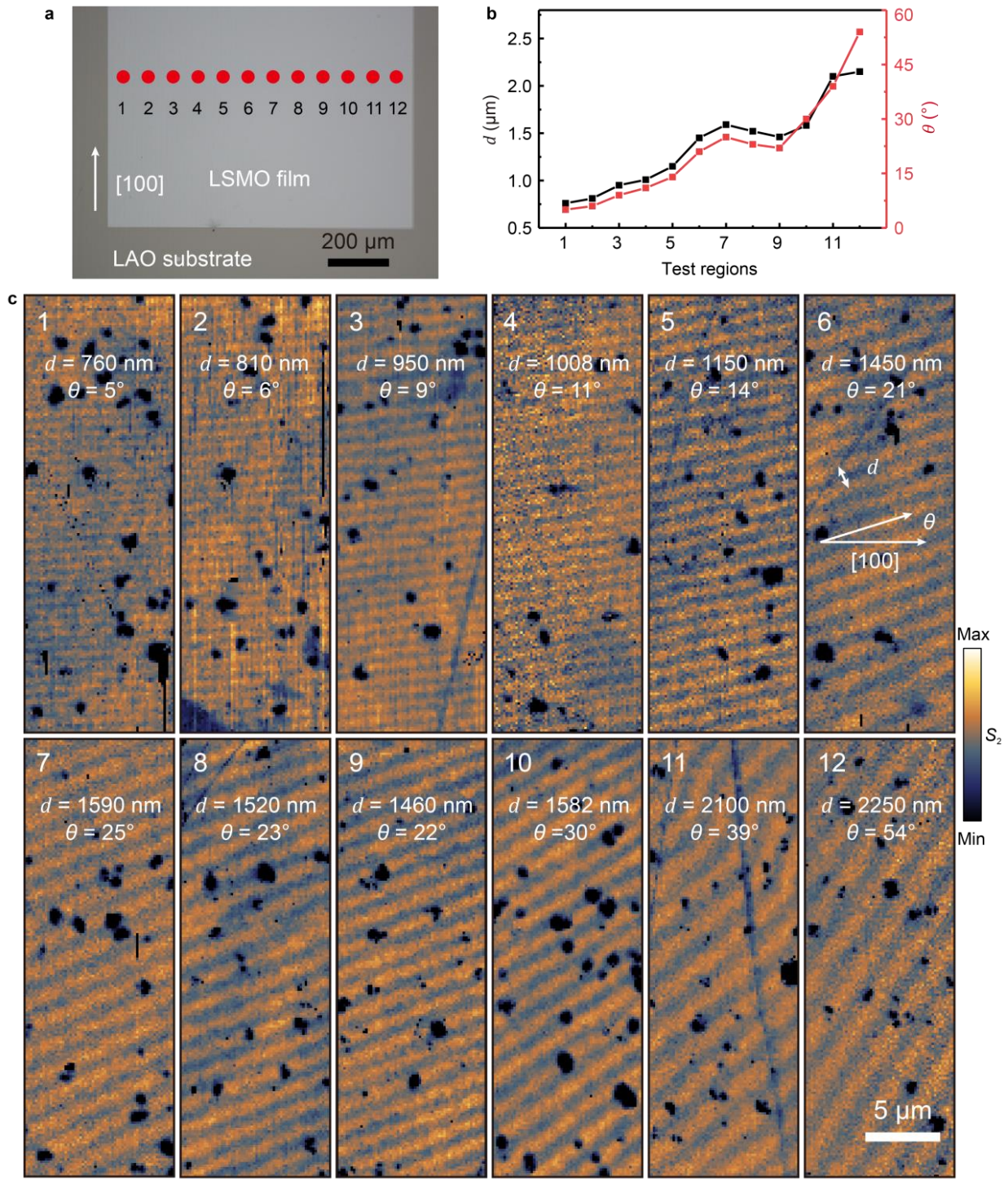

**Extended Data Figure 5 | Variation of electronic moiré patterns across the LSMO film. a**, Optical image of a LSMO thin film on LAO substrate. **b**, The concurrent variations of periodicity $d$ and angle $\theta$ for moiré patterns obtained from different regions (numbered and marked in **a** by red circles). **c**, IR near-field images taken from different regions (1 to 12) marked in **a**.

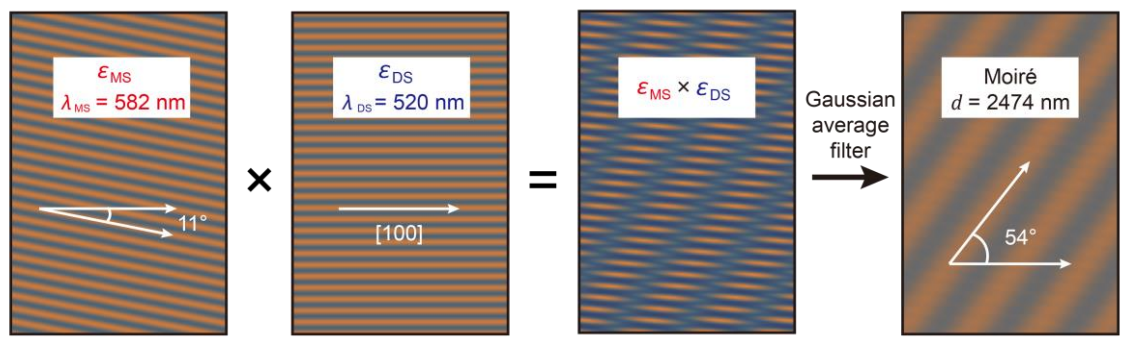

**Extended Data Figure 6 | Simulation of moiré pattern.** The moiré pattern can be simulated by multiplying two periodic striped patterns representing DS and MS and then imposing a Gaussian average filter with radius ~100 nm. The simulation details are described in Supplementary Note 4.

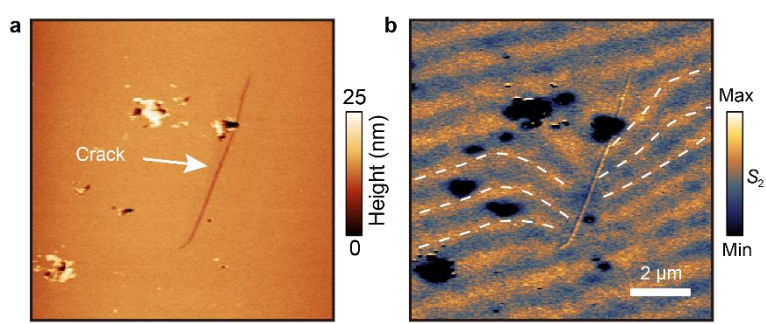

**Extended Data Figure 7 | Influence of the surface crack on moiré pattern geometry. a, b,** AFM and corresponding near-field images taken on a specific region where the moiré patterns are distorted by a surface crack. Distorted moiré patterns are traced by the guiding white dashed lines.

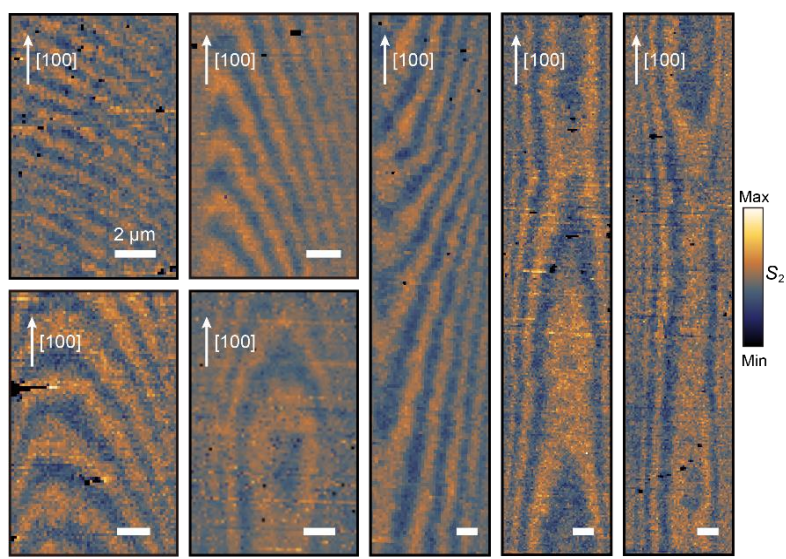

**Extended Data Figure 8 | Different manifestations of moiré patterns observed across the film with vastly distinctive periodicities.**

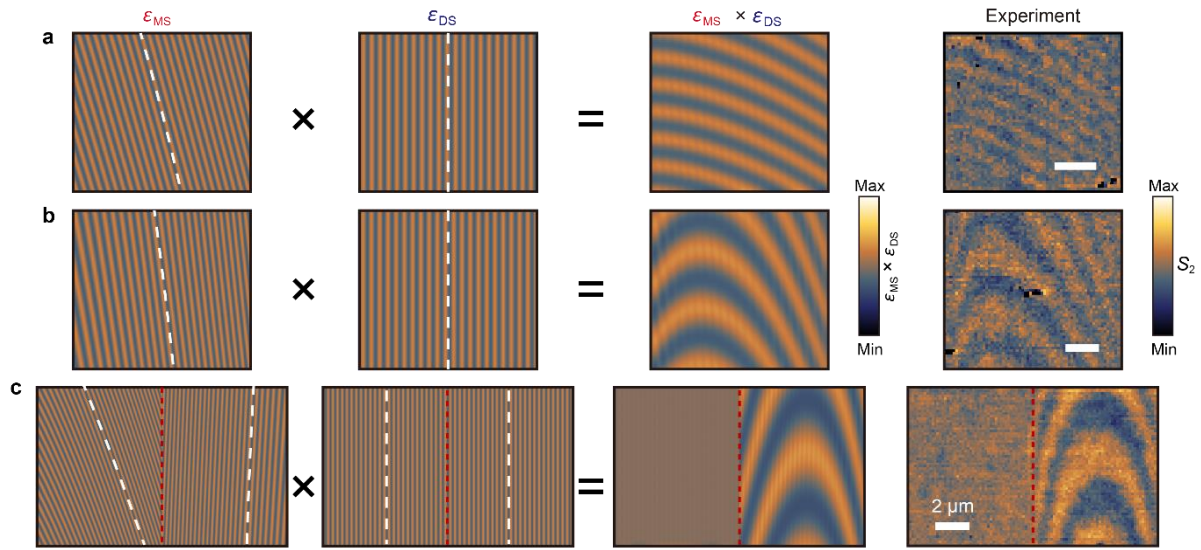

**Extended Data Figure 9 | Simulations of the curved moiré patterns. a-c,** Typical manifestations of curved moiré patterns (right panel) and how they are simulated (left panel). White dashed lines indicate the directions of MS and DS. Curved MS was achieved via adding a spatially varying phase $\phi(y)$ into $\varepsilon_{MS}$ (the MS-induced strain field) during the simulation (see Supplementary Notes 4 and 5). Red dashed lines indicate the LAO twin boundaries.

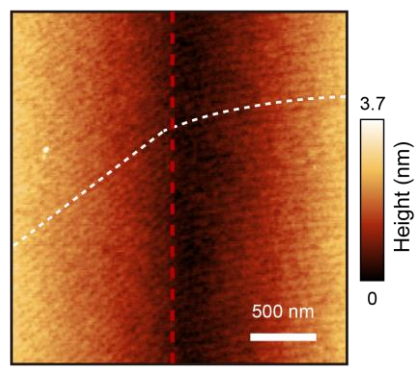

**Extended Data Figure 10 | Diversity of miscut steps in the LAO substrates.** AFM image of a LAO substrate revealing that the miscut steps change their orientations (traced by the white dashed lines) and periodicities across the twin boundary of LAO (indicated by the red dashed line). This is common in commercially available LAO substrates.

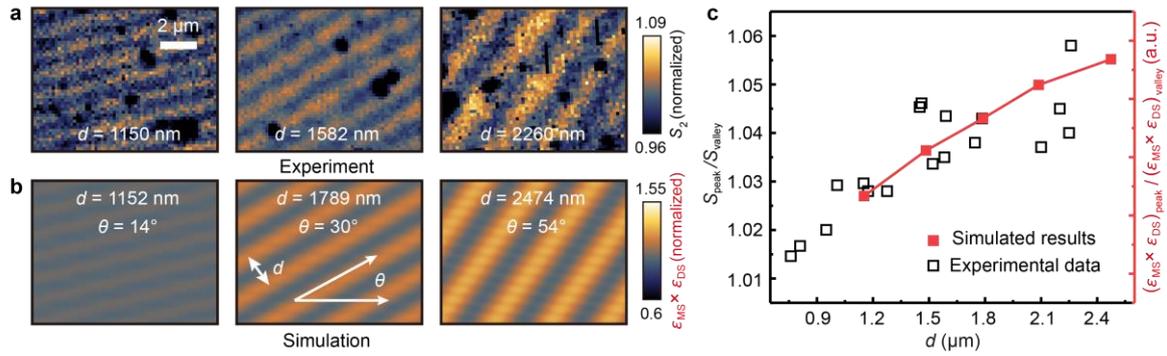

**Extended Data Figure 11 | Further validation of the moiré pattern simulations. a, b,** Typical moiré fringes and corresponding simulations, respectively. The peak and valley contrast of the near-field signal shows an increasing trend with increasing $d$. The data presented here are selected from the main text Figs. 2a, b with the color scale normalized to the dark fringes to better demonstrate their contrast. **c,** The peak and valley ratio of the experimental data (black) and simulated results (red) as a function of $d$. It is clear that the peak/valley contrast is enhanced with increasing moiré pattern periodicity.

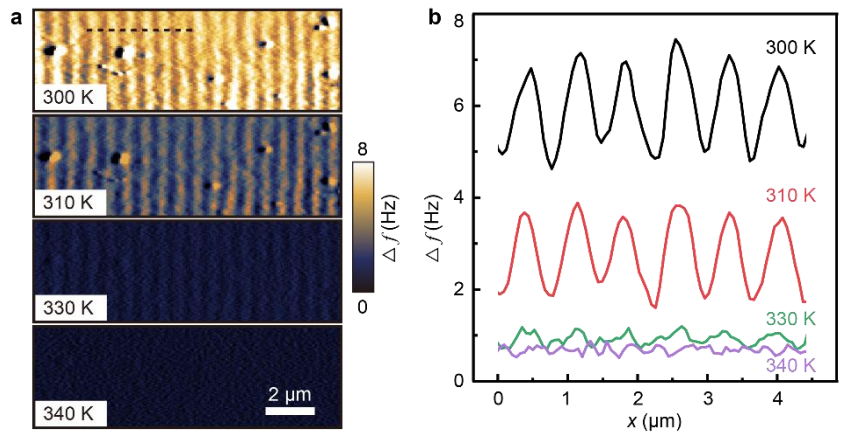

**Extended Data Figure 12 | Temperature-dependent MFM data of moiré patterns. a, b,** MFM images and corresponding line profiles of moiré patterns in a LSMO/LAO sample. The MFM images in **a** are identical to those shown in the main text Fig. 4b, but with the color bar indicating the absolute value of frequency shift. It is clear that the ferromagnetic signal decreases with increasing temperature, and eventually disappears at 340 K, corresponding to a ferromagnetic-paramagnetic transition of LSMO.

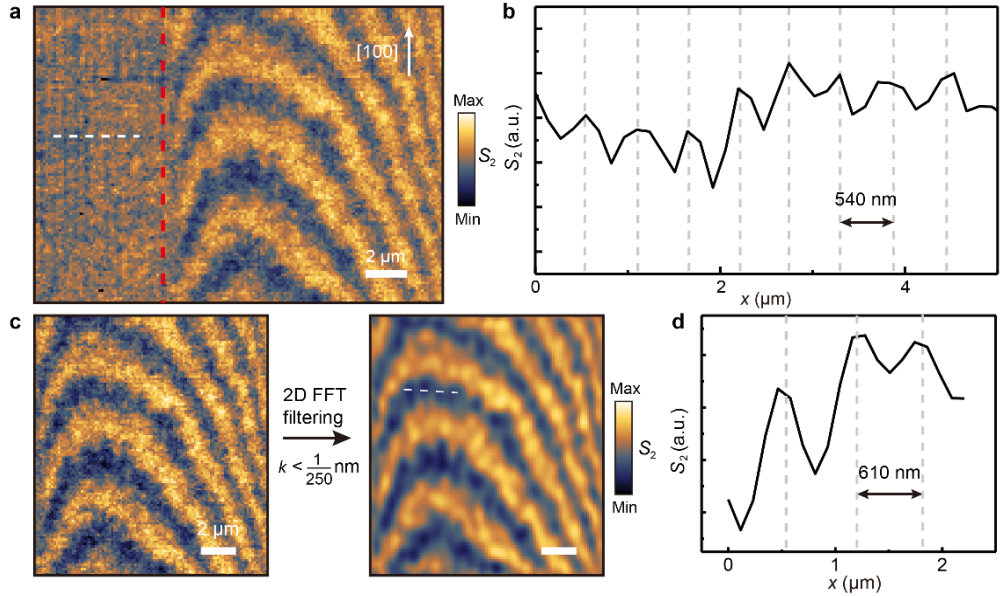

**Extended Data Figure 13 | Evidence of the fine structures. a,** Near-field image taken across the LAO twin boundary (indicated by the red dashed line). On the non-moiré region (left), DS-induced electronic pattern can be faintly identified in the nano-IR contrast. **b,** Line profile of the white dashed line in **a**. The period is ~540 nm, consistent with the OC-SEM image of DS. **c,** Near-field image taken on the moiré region in **a** and filtered with a Fourier filter for $k < \frac{1}{250\,\text{nm}}$. Fine structures of electronic pattern corresponding to MS (or DS) can be observed. **d,** Line profile across the fine structure. The period is ~610 nm, qualitatively consistent with the typical MS (or DS) periodicity.

|  | **LSMO/LAO** | **LSMO/STO** |
|---|---|---|
| Crystal structure (RT) | Rhombohedral/Rhombohedral | Rhombohedral/Cubic |
| Lattice constant (Å) | 3.87/3.79 | 3.87/3.90 |
| Lattice misfit strain | -2.1% | 0.8% |
| Rhombohedral angle | 90.4°/90.1° | 90.4°/90.0° |
| Shear misfit strain | 0.5% | 0.7% |

**Extended Data Table 1 | Comparison of the substrate-induced misfit strain between LSMO/STO and LSMO/LAO.** RT: room temperature. Lattice misfit strain arises from the lattice constant difference, while shear misfit strain is induced by the rhombohedral angle deviation.

## Supplementary Note 1: Twin structures in the LAO substrates

LAO possesses a twin structure at room temperature (Extended Data Table 1). Micron-sized twin structures spontaneously form to compensate the internal strain upon cooling through the phase transition at ~544 °C, at which point the LAO transforms from the cubic phase to a low-symmetry rhombohedral phase[1,2]. Extended Data Fig. 2a shows the optical image of a LAO substrate with the surface partially covered by LSMO thin film. Clearly visible twin structures lie along the single [100] direction of the LAO substrate. Such twin structures further produce identical surface morphology in the epitaxially grown LSMO thin films (main text Fig. 3e). The typical width of the twin structure is of the order of several μm, and the twinning leads to obvious surface corrugation with a relief height up to ~10 nm, as demonstrated in the atomic force microscopy (AFM) image (Extended Data Fig. 2b) and corresponding height profile (Extended Data Fig. 2c).

**Supplementary Note 2:  Rhombohedral domains in the LSMO film on LAO substrate**

The formation of rhombohedral domains has been widely observed in LSMO films grown on cubic $SrTiO_3$ (STO) substrates, which was explained as the relaxation of misfit shear strain[3–7]. For LSMO/STO, the rhombohedral domains can be obtained in thin films with thicknesses ranging from several nm to hundreds of nm[5]. Such a wide range of thicknesses can be attributed to that the lattice misfit strain induced by the lattice constant difference between LSMO and STO is relatively small (Extended Data Table 1), rendering other competing strain relaxation effects, e.g., formation of misfit dislocations, negligible[8]. On the other hand, for LSMO/LAO, the lattice constant mismatch is comparatively large, such that different relaxation mechanisms may compete with each other. As a result, the formation of rhombohedral domains within LSMO films is very sensitive to the growth parameters[9]. In our case the domain formation is only observed within a narrow range of the LSMO thickness around 20 nm.

Extended Data Fig. 3 shows the large-scale OC-SEM image of a 20 nm LSMO thin film on LAO substrate (220 μm × 160 μm), where clear rhombohedral domains running along a uniform direction are evident over the entire measured window. The statistically larger size of the observed rhombohedral domains in our LSMO/LAO thin films (>500 nm) compared to the LSMO/STO case (<100 nm) could be attributed to the relatively small substrate-induced shear misfit strain, as demonstrated in Extended Data Table 1. This is because LAO possesses the same rhombohedral structure as LSMO at room temperature, in contrast to the cubic structure of STO. Furthermore, the direction of the rhombohedral domains in LSMO are aligned along the twin boundary of LAO, as schematically shown in the inset of Extended Data Fig. 3. Such characteristics could be understood as follows: the unidirectional twin structures in the LAO substrates (as shown in Extended Data Fig. 2) break the C4 symmetry, therefore adding a unidirectional restriction on the LSMO rhombohedral domains.

The formation of unidirectional rhombohedral domains results in spatial strain variations in LSMO film on LAO substrate: the LSMO at the domain boundaries experiences less in-plane compressive strain compared to that close to the middle of the domains (main text Fig. 1c). Such periodic strain modulations, named as domain stripes (DS), would in principle generate a striped electronic pattern[7,10]. However, the DS-induced periodic electronic patterns are only faintly observed under s-SNOM imaging (Extended Data Figs. 13a, b). This is likely because DS alone are not sufficient to induce significant strain and therefore optical conductivity modulation that could be detected under s-SNOM imaging. The moiré pattern formation magnifies the local strain by adding additional effect induced by MS. Therefore, the moiré pattern observed in s-SNOM better reveals the "hidden" electronic order induced by DS.

**Supplementary Note 3: Miscut stripes (MS) and induced electronic pattern**

The miscut steps in the LAO substrate cause periodic strain modulations in LSMO. For example, LSMO close to the step edges is expected to experience less in-plane compression (main text Fig. 1e). Analogous to the DS, we name such miscut steps-induced periodic strain in LSMO as miscut stripes (MS). As demonstrated below, the MS could in turn lead to spatial conductivity modulation, as the LSMO under less in-plane compression normally corresponds to stronger metallicity and higher optical response[11–13].

As discussed above, the existence of rhombohedral domains in LSMO/LAO thin films is quite sensitive to the growth parameters. For some certain samples no visible rhombohedral domain could be detected using OC-SEM. The absence of DS thus offers the opportunity to detect pure MS-induced electronic patterns. We conducted the AFM and s-SNOM measurements in such samples, and the typical results are shown in Extended Data Fig. 4. Miscut steps with ~500 nm periodicity are clearly identified by AFM imaging (Extended Data Fig. 4a), while simultaneously acquired near-field contrast coincides with the sample topography (Extended Data Fig. 4b). Note that the metallicity and the surface height are positively correlated (Extended Data Fig. 4c), which rules out the possibility of false signal contrast that reportedly shows negative correlation with topography[14].

**Supplementary Note 4: Moiré patterns simulations**

To simulate the moiré pattern in main text Fig. 2a, we first define two scalar sinusoidal strain fields corresponding to the in-plane strain caused by DS and MS, respectively. Namely, $\varepsilon_{DS}(\boldsymbol{k}_{DS}, \boldsymbol{x}) = A_1 \sin(\boldsymbol{k}_{DS} \cdot \boldsymbol{x}) + B_1$ and $\varepsilon_{MS}(\boldsymbol{k}_{MS}, \boldsymbol{x}) = A_2 \sin(\boldsymbol{k}_{MS} \cdot \boldsymbol{x}) + B_2$, where $\varepsilon_{DS}$ and $\varepsilon_{MS}$ are the microscopic strain fields induced by DS and MS, respectively. $\boldsymbol{k}_{DS}$ and $\boldsymbol{k}_{MS}$ are the wavevectors of DS and MS, respectively. To model and understand the strain dependent near-field fringes, we make the assumptions that: (i) fringes are a direct reporter of the optical conductivity[15,16]; (ii) the conductivity depends nonlinearly on strain[12,17].

We have:

$$\sigma \propto \sigma_0 + A(\varepsilon_{DS} + \varepsilon_{MS}) + B(\varepsilon_{DS} + \varepsilon_{MS})^2$$

$$= \sigma_0 + A(\varepsilon_{DS} + \varepsilon_{MS}) + B(\varepsilon_{DS}^2 + \varepsilon_{MS}^2 + \boldsymbol{2\varepsilon_{DS}} \cdot \boldsymbol{\varepsilon_{MS}}). \tag{1}$$

The factors $A$ and $B$ are first- and second-order Taylor coefficients, respectively. We were then able to reproduce the key trends in the evolution of the fringe patterns in main text Fig. 2a, finding that the fringes are the direct manifestation of the cross term $\boldsymbol{\varepsilon}_{DS} \cdot \boldsymbol{\varepsilon}_{MS}$. The existence of the large cross term $\varepsilon_{DS} \cdot \varepsilon_{MS}$ attests to the previous observation of the strain-induced $T_C$ modulation in LSMO[11,12,17,18], where $T_C \propto 1 - a\varepsilon_{xx} - b\varepsilon_{xx}^2$, assuming the unit cell volume is preserved. Here $\varepsilon_{xx}$ is the in-plane strain, $a = \left(\frac{1}{T_C}\right)(dT_C/d\varepsilon_{xx})$ and $b = \left(\frac{1}{T_C}\right)(d^2T_C/d\varepsilon_{xx}^2)$. The quadratic coefficient $b$ has been experimentally demonstrated to be orders of magnitude larger than the linear coefficient $a$[13], consistent with our observation. The large magnitude of the nonlinear coefficients ($B$ and $b$) originates from the cubic (Oh) point symmetry of Mn in the ideal cubic perovskite structure as well as the large Jahn-Teller effect due to the strong electron-lattice coupling in LSMO[19].

The validity of using $\varepsilon_{DS} \cdot \varepsilon_{MS}$ for the simulation can be further understood in a Landau free energy scheme. According to the equation (1), periodic structures in the $\sigma$ map can occur with six possible periodicities, represented by their wavevectors: $\boldsymbol{k}_{DS}, \boldsymbol{k}_{MS}, 2\boldsymbol{k}_{DS}, 2\boldsymbol{k}_{MS}, \boldsymbol{k}_{DS} + \boldsymbol{k}_{MS}$, and $\boldsymbol{k}_{DS} - \boldsymbol{k}_{MS}$, with $\boldsymbol{k}_{DS}$ originating from the term $\varepsilon_{DS}$, $\boldsymbol{k}_{MS}$ from $\varepsilon_{MS}$, $2\boldsymbol{k}_{DS}$ from $\varepsilon_{DS}^2$, $2\boldsymbol{k}_{MS}$ from $\varepsilon_{MS}^2$, $\boldsymbol{k}_{DS} + \boldsymbol{k}_{MS}$ and $\boldsymbol{k}_{DS} - \boldsymbol{k}_{MS}$ from the cross term $\varepsilon_{DS} \cdot \varepsilon_{MS}$, respectively. In a Landau free energy scheme[20], the free energy $F$ as a function of local order parameter $\varphi$ (e.g. magnetization or lattice distortion) can be written as the sum of the domain wall energy (gradient term) and the single-well potential, i.e. $F = \frac{1}{2}\kappa|\nabla\varphi|^2 + \alpha'(T - T_C)\varphi^2 + \gamma\varphi^4$, where $\kappa$ is the "stiffness" coefficient, $\alpha'$ and $\gamma$ are second- and forth-order coefficients for the potential well. In the case when $\kappa$ is sufficiently large, fast varying fringes ($\boldsymbol{k}_{DS}, \boldsymbol{k}_{MS}, 2\boldsymbol{k}_{DS}, 2\boldsymbol{k}_{MS}$, and $\boldsymbol{k}_{DS} + \boldsymbol{k}_{MS}$) are not energetically favorable and the slowly varying fringes ($\boldsymbol{k}_{DS} - \boldsymbol{k}_{MS}$) are more likely to survive. Thus, the effects produced by these fast-varying fringes are expected to smooth out spontaneously, which leads to the experimental manifestation primarily depending on the $\boldsymbol{k}_{DS} - \boldsymbol{k}_{MS}$ term originating from $\varepsilon_{DS} \cdot \varepsilon_{MS}$.

The lower half of main text Fig. 2b was simulated by calculating $\varepsilon_{DS} \cdot \varepsilon_{MS}$, followed by imposing a local Gaussian average filter with radius ~100 nm (Extended Data Fig. 6). The later smoothing processing also accounts for the reduced contrast of high spatial frequency term $\boldsymbol{k}_{DS} + \boldsymbol{k}_{MS}$. Using this method, we can perfectly reproduce the textures of various types of moiré fringes, including their periodicities, curvatures, and peak-to-valley contrasts.

## Supplementary Note 5: Simulation of Curved moiré patterns

The variable miscut steps in the LAO substrate offers a unique opportunity to realize a diverse range of electronic moiré textures with different configurations. In main text Fig. 3 and Extended Data Fig. 8, different types of electronic moiré patterns are presented. As schematically shown in Extended Data Fig. 9, these seemingly complicated patterns could be well reproduced by $\varepsilon_{DS}(\boldsymbol{k}_{DS}, \boldsymbol{x}) \cdot \varepsilon_{MS}(\boldsymbol{k}_{MS}, \boldsymbol{x})$. Differing from the straight MS utilized for the simulation of straight moiré patterns in Extended Data Fig. 6, here MS with locally varying curvature, periodicity, or direction are used to obtain curved moiré patterns. As a simple demonstration we set $\varepsilon_{MS} = A_2 \sin(\boldsymbol{k}_{MS} \cdot \boldsymbol{x} + \phi(y)) + B_2$ during our simulations, where $\phi(y) = k(\frac{y}{L})^l$. $L$ is the simulation box length, and $k$ and $l$ determine how fast $\phi$ varies. Numerically, in our simulations we keep $l = 1.65$ constant and vary $k$ to fit the experimental data.

## Supplementary Note 6: Fine structures of moiré patterns

Note that in some near-field images, the fine structures (with a wavevector of $\boldsymbol{k}_{DS}$ or $\boldsymbol{k}_{MS}$) can be faintly observed (Extended Data Fig. 13c). With a 2D Fast Fourier Transform (FFT) filter, the signals possessing high spatial frequencies of $k > \frac{1}{250\,nm}$ can be effectively filtered out. An image with sharper contrast focusing on the relatively large periodicities can be obtained (right panel of extended Data Fig. 13c), in which the fine structures of the curved moiré patterns are evident. From the corresponding line profiles in the filtered image, a periodic oscillation with periodicity of ~610 nm is revealed (extended Data Fig. 13d), which is consistent with the typical periodicity of the MS (or DS). This observation demonstrates that both the $\boldsymbol{k}_{DS}$ and the $\boldsymbol{k}_{MS}$ terms can be potentially observed in the moiré pattern region, but with a much lower contrast than that of the moiré pattern. Lastly, we remark that the geometry of the moiré fringes was observed to be highly sensitive to surface cracks or defects. As shown in Extended Data Fig. 7, surface cracks can result in locally curved moiré pattern. This is likely because the DS adapts differently in the vicinity of the cracks, producing distinct changes in the moiré pattern.